\documentclass[journal=jpclcd,manuscript=letter]{achemso}

\usepackage[version=3]{mhchem} 

\usepackage[usenames,dvipsnames]{xcolor}

\usepackage[defaultcolor=blue]{changes}
\usepackage{nameref}

\author{Lina Fransén}
\affiliation{Nantes Université, CNRS, CEISAM UMR 6230, F-44000 Nantes,France}
\author{Sandra Gómez}
\affiliation{Department of Chemistry,
Universidad Autónoma de Madrid,
C. Francisco Tomás y Valiente, 7, 28049 Madrid}
\author{Morgane Vacher}
\affiliation{Nantes Université, CNRS, CEISAM UMR 6230, F-44000 Nantes,France}
\email{morgane.vacher@univ-nantes.fr}

\title[An \textsf{achemso} demo]
  {Attochemical Control of Nuclear Motion Despite Fast Electronic Decoherence}

\begin{document}
\begin{tocentry}

\includegraphics[]{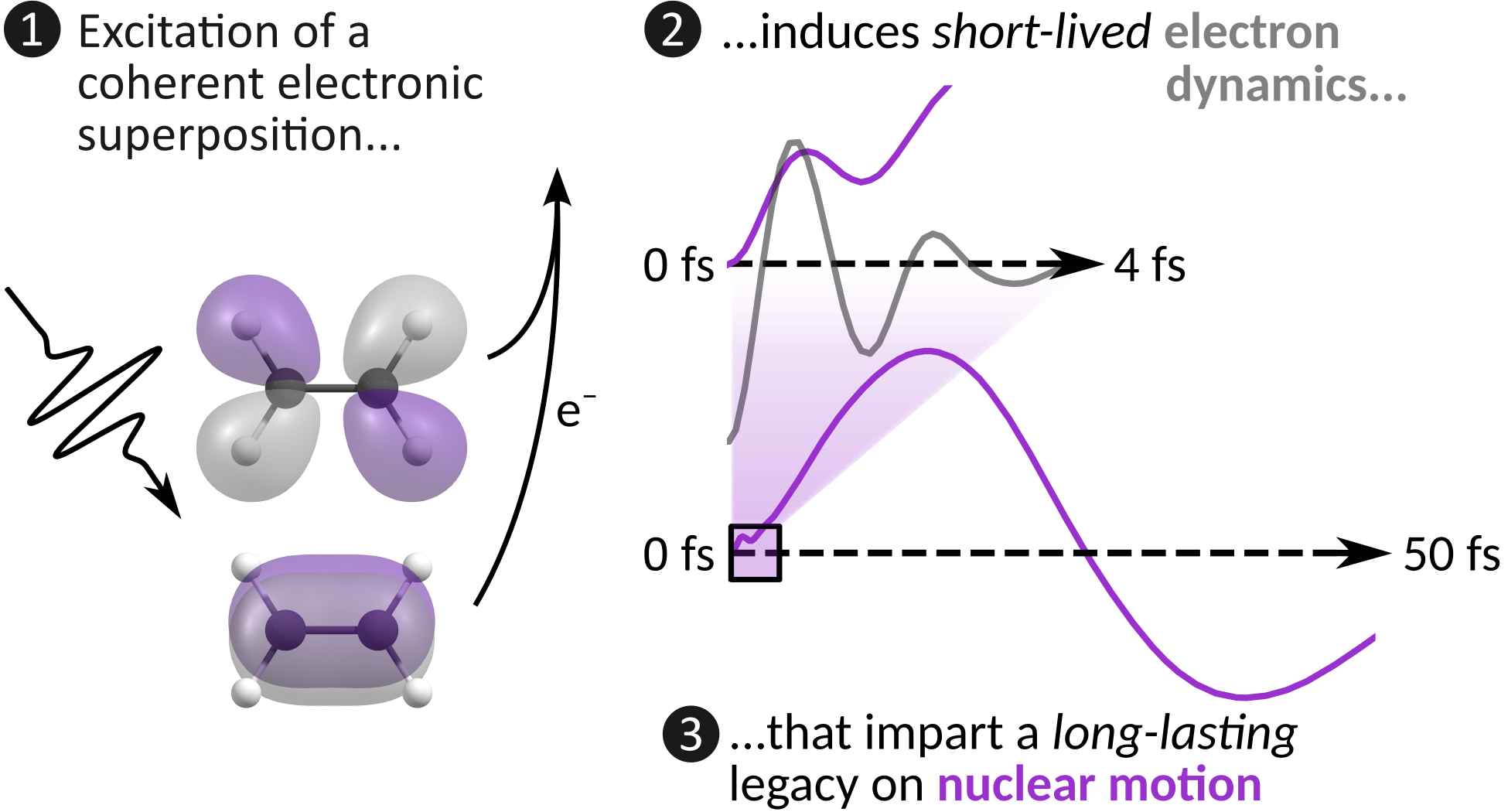}






\end{tocentry}

\begin{abstract}

\noindent Short-in-time, broad-in-energy atto- or few-femtosecond pulses can excite coherent superpositions of several electronic states in molecules. This results in ultrafast charge oscillations known as charge migration. A key open question in the emerging field of attochemistry is whether these electron dynamics, which due to decoherence often last only for a few femtoseconds, can influence longer-timescale nuclear rearrangements. Herein, we address this question through full-dimensional quantum dynamics simulations of the coupled electron-nuclear dynamics initiated by ionization and coherent excitation of ethylene. The simulations on this prototype organic chromophore predict electronic coherences with half-lives of less than one femtosecond. Despite their brevity, these electronic coherences induce vibrational coherences along the derivative coupling vectors that persist for at least 50\,fs. These results suggest that short-lived electronic coherences can impart long-lasting legacies on nuclear motion, a finding of potential importance to the interpretation of attosecond experiments and the development of strategies for attochemical control.
\end{abstract}


The experimental realization of attosecond pulses at the beginning of the 21$^{\ce{st}}$ century\cite{Paul_Toma_Breger_Mullot_Augé_Balcou_Muller_Agostini_2001, Hentschel_Kienberger_Spielmann_Reider_Milosevic_Brabec_Corkum_Heinzmann_Drescher_Krausz_2001} has enabled the excitation of coherent superpositions of electronic states in photoionized molecules. This results in a purely electronic motion, known as charge migration\cite{Cederbaum_Zobeley_1999}, where the hole density oscillates along the molecular backbone with frequencies defined by the energy gaps between the involved electronic states. Control of the motion of the electrons---the particles responsible for chemical bonds---through the excitation of tailored electronic wave packets may open up the possibility of steering chemical reactivity toward a desired outcome. This is the paradigm of attochemistry,\cite{Lepine_Ivanov_Vrakking_2014, Merritt_Jacquemin_Vacher_2021, Calegari_Martin_2023} which offers an alternative to femtochemical control strategies\cite{Shapiro_Brumer_2001, Brumer_Shapiro_1989, Charron_Giusti-Suzor_Meis_1995, Tannor_Kosloff_Rice_1986, Assion_Baumert_Bergt_Brixner_Kiefer_Seyfried_Strehle_Gerber_1998, Brixner_Damrauer_Niklaus_Gerber_2001} that seek to steer reactivity by instead controlling the nuclear motion through the excitation of tailored nuclear wave packets. However, electronic coherences in molecules often only last for a few femtoseconds.\cite{Vacher_Bearpark_Robb_Malhado_2017, Arnold_Vendrell_Welsch_Santra_2018, Scheidegger_Vaníček_Golubev_2022, Arnold_Vendrell_Santra_2017, Jia_Manz_Yang_2019, Despré_Golubev_Kuleff_2018} This is widely believed to pose a challenge for controlling chemical transformations attochemically,\cite{Nisoli_2024, Vester_Despré_Kuleff_2023, Despré_Golubev_Kuleff_2018, Merritt_Jacquemin_Vacher_2021, Alexander-2023} as these unfold on much longer timescales.

This letter demonstrates theoretically that short-lived electronic coherences can impart long-lasting legacies on the nuclear motion. This is done through quantum dynamics simulations of ethylene upon ionization to its three lowest-lying cationic electronic states ($\tilde{\ce{X}}$, $\tilde{\ce{A}}$, and $\tilde{\ce{B}}$). Ethylene cation is the simplest organic $\pi$ radical and has previously been extensively studied experimentally using attosecond technology\cite{Tilborg-2009, Ludwig-2016, Zinchenko-2021, Lucchini-2022, Vacher_Albertani_Jenkins_Polyak_Bearpark_Robb_2016, Lucchini_Cardosa-Gutierrez_Murari_Frassetto_Poletto_Nisoli_Remacle_2025} and theoretically through non-adiabatic dynamics simulations\cite{Joalland-2014, Ludwig-2016, Zinchenko-2021, Vacher-2022, Fransén_Tran_Nandi_Vacher_2024, Tran-2025, Lucchini_Cardosa-Gutierrez_Murari_Frassetto_Poletto_Nisoli_Remacle_2025}. 

\begin{figure}[H]
    \centering
    \includegraphics[]{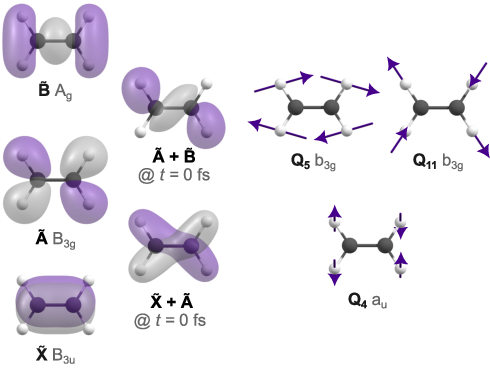}
    \caption{Left: Isodensity plots of the hole (i.e., singly occupied) molecular orbitals that characterize the individual ${\tilde{\ce{X}}}$, ${\tilde{\ce{A}}}$, and ${\tilde{\ce{B}}}$ states. Middle: Schematic representation of the holes associated with the $\tilde{\ce{X}}$+$\tilde{
    \ce{A}}$ and $\tilde{\ce{A}}$+$\tilde{\ce{B}}$ coherent superpositions. Right: The vibrational normal modes along which the inter-state gradients (defined in Equation \ref{eq:NuclearGrad}) are non-zero for the two coherent superpositions.}
    \label{fig:Fig0}
\end{figure}

As an illustrative case, we focus on in-phase coherent superpositions of pairs of equally weighted cationic diabatic electronic states. The conclusions are, however, expected to be generalizable to other superpositions; this is supported by results for three-state superpositions (Section S8 of the Supporting Information). The initial wavefunctions read

\begin{equation}
\label{eq:InitSup}
\begin{split}
        \Psi(\boldsymbol{r}, \boldsymbol{Q}, t = 0) &= \left[c_i\psi_i  (\boldsymbol{r}) + c_j\psi_j (\boldsymbol{r})e^{i\varphi} \right]\chi_0(\boldsymbol{Q}, t = 0) \\
    &= \Psi_{\ce{el}}(\boldsymbol{r})\chi_0(\boldsymbol{Q}, t=0)
\end{split}
\end{equation}

\noindent where $c_i = c_j = 1/\sqrt{2}$ are the coefficients of the diabatic electronic states $\psi_i$ and $\psi_j$, $\varphi = 0$ is the initial relative phase between the electronic states, and $\chi_0$ is the ground state vibrational wave function of neutral ethylene. Two superpositions are considered: one comprising the $\tilde{\ce{X}}$ and $\tilde{\ce{A}}$ states, and another involving the $\tilde{\ce{A}}$ and $\tilde{\ce{B}}$ states. 
Figure \ref{fig:Fig0} displays the hole (i.e., singly occupied) orbitals that characterize the individual $\tilde{\ce{X}}$, $\tilde{\ce{A}}$, and $\tilde{\ce{B}}$ states, along with schematic representations of the holes created upon ionization and excitations of the $\tilde{\ce{X}}$+$\tilde{\ce{A}}$ and $\tilde{\ce{A}}$+$\tilde{\ce{B}}$ coherent superpositions. Due to interference, the hole densities of the coherent superpositions are time-dependent and differ from the incoherent averages of the hole densities of the individual states. 

Here, we note in passing that one-photon ionization to the $\tilde{\ce{X}}$ + $\tilde{\ce{A}}$ superposition would, within the electric dipole approximation, result in a fully incoherent state, as the hole orbitals---and thus the photoelectrons---associated with the $\tilde{\ce{X}}$ and $\tilde{\ce{A}}$ channels have opposite parities. A coherent superposition of the $\tilde{\ce{X}}$ and $\tilde{\ce{A}}$ states may, nevertheless, be realizable using a sequence of pulses (cf. Ref.\citenum{Sansone_Kelkensberg_Pérez_Torres_Morales_Kling_Siu_Ghafur_Johnsson_Swoboda_Benedetti_et_al._2010}, where a coherent superposition of the $\Sigma_{\ce{g}}^+$ and $\Sigma_{\ce{u}}^+$ states of \ce{D2+} was generated experimentally).

A fully quantum mechanical treatment of the coupled electron-nuclear dynamics is required to adequately simulate the nuclear motion induced by coherent superpositions of electronic states.\cite{Tran_Ferté_Vacher_2024} Here, we employ one of the most accurate approaches--the multilayer multiconfiguration time-dependent Hartree (ML-MCTDH)\cite{Wang_Thoss_2003} method--combined with a vibronic coupling Hamiltonian\cite{Köppel_Domcke_Cederbaum_1984} parametrized along the 12 vibrational normal modes of ethylene. The simulations are performed in full nuclear dimensionality to adequately capture the electronic decoherence: previous work has shown that decoherence can result from the interplay of many vibrational modes.\cite{Arnold_Vendrell_Santra_2017}. 

Vibronic coupling models have previously been used to calculate photoelectron spectra of ethylene.\cite{Köppel_Domcke_Cederbaum_Niessen_1978, Köppel_Cederbaum_Domcke_1982, Hazra_Nooijen_2005_1, Hazra_Nooijen_2005_2} In the present work, the vibronic coupling Hamiltonian, which is diabatic by ansatz, reads

\begin{equation}
\label{eq:vibronic_coupling_ansatz}
    \boldsymbol{H} = \left( \hat{T}_{\ce{N}}+ V_0 \right) \boldsymbol{1} + \boldsymbol{W}
\end{equation}

\noindent where $\hat{T}_{\ce{N}}$ is the nuclear kinetic energy operator, $V_0$ is the zero-order diabatic potential, and $\boldsymbol{W}$ is the diabatic potential energy matrix. The latter is expanded in a Taylor series around the equilibrium geometry of neutral ethylene in the basis of mass-frequency scaled normal mode coordinates. Its on- and off-diagonal elements read

\begin{equation}
        W_{ii} (\boldsymbol{Q}) = E_i + \sum_{\alpha}^{f}\kappa_{\alpha}^{(i)}Q_\alpha + \sum_{\alpha}^{f}\frac{1}{2}\gamma_{\alpha,\alpha}^{(i)}Q_\alpha^2 
\end{equation}

\begin{equation}
            W_{ij} (\boldsymbol{Q}) = \sum_{\alpha}^{f}\lambda_{\alpha}^{(i,j)}Q_\alpha
\end{equation}
\noindent where $E_i$ is the vertical ionization energy of the cationic electronic state $\psi_i$, $\kappa_{\alpha}^{(i)}$ and $\gamma_{\alpha,\alpha}^{(i)}$ are the linear and quadratic intra-state coupling constants on state $\psi_i$ along normal mode $Q_{\alpha}$, and $\lambda_{\alpha}^{(i,j)}$ is the linear inter-state coupling constant between states $\psi_i$ and $\psi_j$ along mode $Q_{\alpha}$. The effect of including bilinear intra-state coupling constants ($\gamma^{(i)}_{\alpha,\beta}$, $\alpha\neq\beta$) is discussed in Section S6 of the Supporting Information. The parameters entering the model Hamiltonian were obtained from a least-square fit to SA5-CASSCF(11e,12o) energies. Details of the active space are provided in Section S1 of the Supporting Information. The \textit{ab initio} energies were calculated with OpenMolcas\cite{Open-Molcas-2019, Open-Molcas-2023, v22.10} using the ANO-RCC-VDZP\cite{Roos-2004} basis set and the atomic compact Cholesky decomposition.\cite{Aquilante-2009} The fitting,  performed with the VCHam programs in the Quantics software,\cite{Worth_2020} was facilitated by the high point group symmetry (\ce{D_{{$2h$}}}) of ethylene, as the following selection rules limit the number of non-zero parameters:

\begin{subequations}
\begin{align}
    \kappa_{\alpha}^{(i)} \neq 0: \Gamma_{\alpha} \supset \Gamma_{\mathrm{A}} \\
    \lambda_{\alpha}^{(i,j)} \neq 0: \Gamma_{\alpha} \otimes \Gamma_i \otimes \Gamma_j \supset \Gamma_{\mathrm{A}}
\end{align}
\label{eq:Symmetry}
\end{subequations}

\noindent where $\Gamma_{\ce{A}}$ is the totally symmetric irreducible representation of the point group, $\Gamma_{\alpha}$ is a vibrational symmetry, and $\Gamma_i$ and $\Gamma_j$ are electronic state symmetries. The quadratic intra-state coupling constants are symmetry-allowed for all vibrational modes on all electronic states ($\gamma_{\alpha,\alpha}^{(i)}\neq 0: \Gamma_{\alpha}\otimes\Gamma_{\alpha}\supset \Gamma_{\ce{A}}$). Further details on the model Hamiltonian are reported in Section S1 of the Supporting Information, and the robustness of the results with respect to the model parameters is discussed in Section S7. Details on the ML-MCTDH simulations, which were performed with Quantics\cite{Worth_2020}, using (where possible) an adaptive number of single particle functions,\cite{Mendive-Tapia_Meyer_2020} are provided in Section S2.

The electronic population decays obtained using the model Hamiltonian in MCTDH simulations initiated on the individual ${\tilde{\ce{A}}}$ and ${\tilde{\ce{B}}}$ states are in good agreement with results from previous on-the-fly surface hopping simulations\cite{Fransén_Tran_Nandi_Vacher_2024} (Section S3 of the Supporting Information). For a fair comparison, the electronic populations from the MCTDH simulations were transformed from the diabatic to the adiabatic representation\cite{Coonjobeeharry_Spinlove_Sanz_Sapunar_Došlić_Worth_2022, Manthe_1996}. Moreover, the time scale of the electronic relaxation following ionization to the individual $\tilde{\ce{A}}$ state is consistent with experimental results obtained through attosecond transient absorption spectroscopy.\cite{Zinchenko-2021} These agreements with previous results provide validation for the present model Hamiltonian. We acknowledge, however, that the model Hamiltonian is limited to describing the dynamics during the first $\sim$50\,fs after ionization, before the onset of large-amplitude nuclear motion. This and other limitations are discussed in Section S3 of the Supporting Information.

\begin{figure}[h!]
    \centering
    \includegraphics[]{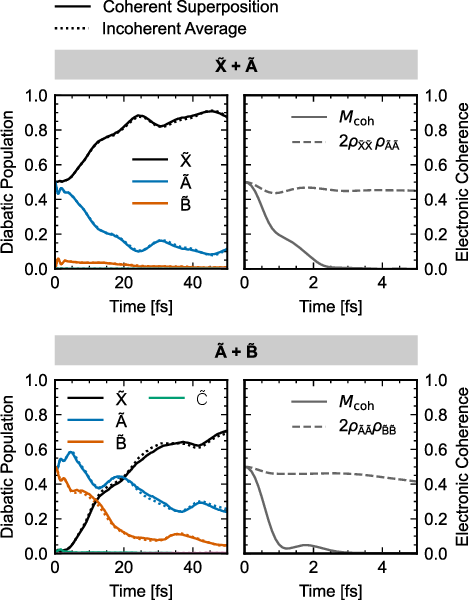}
    \caption{
    Left: Time evolutions of the diabatic electronic populations following excitations of the ${\tilde{\ce{X}}}$+${\tilde{\ce{A}}}$ (top) and the ${\tilde{\ce{A}}}$+${\tilde{\ce{B}}}$ (bottom) equally weighted in-phase coherent superpositions. The dotted lines represent the incoherent averages of the population dynamics obtained following excitations of the relevant individual electronic states.  
    Right: Degrees of electronic coherence (solid lines) and their upper bounds based on the electronic populations (dashed lines) following excitations of the $\tilde{\ce{X}}$+$\tilde{\ce{A}}$ (top) and $\tilde{\ce{A}}$+$\tilde{\ce{B}}$ (bottom) coherent superpositions. Note that the x-axis ranges for the right-hand side plots are ten times shorter than for the left plots.}
    \label{fig:ElPop_ElCoh}
\end{figure}

The time evolutions of the diabatic electronic populations following excitations of the $\tilde{\ce{X}}$+$\tilde{\ce{A}}$ and $\tilde{\ce{A}}$+$\tilde{\ce{B}}$ superpositions are reported in the left-hand side of Figure \ref{fig:ElPop_ElCoh}. The results for the coherent superpositions are compared with incoherent averages of simulation results obtained upon excitations of the individual $\tilde{\ce{X}}$, $\tilde{\ce{A}}$, and $\tilde{\ce{B}}$ states. Quantum coherences between the initially populated states do not significantly influence the population decays, as evidenced by the close agreement between the results for the coherent superpositions and the incoherent averages. 

Electronic state population transfer, dephasing due to the nuclear wave packet width, and loss of overlap between nuclear wave packets associated with different electronic states can lead to electronic decoherence.\cite{Fiete_Heller_2003, Vacher_Albertani_Jenkins_Polyak_Bearpark_Robb_2016} The degree of electronic coherence can be measured using the electronic purity, Tr($\rho^2$), where Tr denotes the trace and $\rho$ is the reduced density matrix. The elements of the reduced density matrix read

\begin{equation}
\begin{split}
    \rho_{ij} = \int d\boldsymbol{Q} \langle{\psi_i(\boldsymbol{r})}|{\Psi(\boldsymbol{r}, \boldsymbol{Q}, t)}\rangle \langle{\Psi(\boldsymbol{r}, \boldsymbol{Q}, t)|\psi_j(\boldsymbol{r})\rangle} \\ = \int d\boldsymbol{Q} \chi_{i}(\boldsymbol{Q}, t) \chi_{j}^* (\boldsymbol{Q}, t)
\end{split}
\end{equation}

\noindent where the on-diagonal terms $\rho_{ii}$ represent the electronic populations and the off-diagonal ones $\rho_{ij}$ correspond to the electronic coherences.
The right-hand side of Figure \ref{fig:ElPop_ElCoh} reports the term $M_{\ce{coh}}=2\rho_{ij}\rho_{ji}$\cite{Fiete_Heller_2003, Vacher_Albertani_Jenkins_Polyak_Bearpark_Robb_2016}, which quantifies the electronic coherence by gathering the off-diagonal components of Tr($\rho^2$). 
One can show that $M_{\ce{coh}}\leq 2\rho_{ii}\rho_{jj}$.\cite{Vacher_Albertani_Jenkins_Polyak_Bearpark_Robb_2016} At $t=0$\,fs,  $M_{\ce{coh}}$=0.5, the theoretical maximum for equally weighted two-state superpositions.  
Decoherence then occurs rapidly, with $M_{\ce{coh}}$ dropping to half of its initial value after $\sim$0.8\,fs and $\sim$0.7\,fs for the $\tilde{\ce{X}}$+$\tilde{\ce{A}}$ and the $\tilde{\ce{A}}$+$\tilde{\ce{B}}$ superpositions, respectively. Similar decoherence time scales have previously been reported in theoretical studies on other molecules.\cite{Vacher_Bearpark_Robb_Malhado_2017, Arnold_Vendrell_Welsch_Santra_2018, Scheidegger_Vaníček_Golubev_2022, Arnold_Vendrell_Santra_2017, Jia_Manz_Yang_2019, Despré_Golubev_Kuleff_2018} In the present case, the rapid electronic decoherences are not primarily due to electronic state population transfers, as the $M_{\ce{coh}}$ values decay much more quickly than the $2\rho_{ii}\rho_{jj}$ values. Instead, the decoherence is mainly caused by dephasing due to the nuclear wave packet width and loss of overlap between nuclear wave packets associated with different electronic states. Given the rapidity of the decoherence, it is reasonable to assume that the former mechanism---dephasing---dominates. Unlike the latter, this mechanism does not require nuclear wave packets to move; it can occur due to the width of the nuclear wave packets instantaneously after ionization. Dephasing has been identified as the main cause of decoherence in previous fully quantum mechanical dynamics studies on other molecules.\cite{Vacher_Bearpark_Robb_Malhado_2017, Arnold_Vendrell_Santra_2017}

Having established that the electronic coherences are short-lived and do not affect the electronic population decays, we now investigate their impact on the nuclear motion. The nuclear energy gradient at $Q=0$ and $t=0$\,fs along normal mode $Q_\alpha$ for a coherent superposition described by Equation \ref{eq:InitSup} is given by\cite{Meisner_Vacher_Bearpark_Robb_2015}

\begin{equation}
\label{eq:NuclearGrad}
\begin{split}
    \frac{\partial}{\partial Q_{\alpha}}\langle \Psi_{\ce{el}}(\boldsymbol{r}) |\hat{H}_{\mathrm{el}} | \Psi_{\ce{el}}(\boldsymbol{r}) \rangle = {|c_i|}^2 \kappa_{\alpha}^{(i)} \\ + {|c_j|}^2 \kappa_{\alpha}^{(j)} + 2|c_i||c_j|\mathrm{cos}(\varphi)\lambda_{\alpha}^{(i,j)}
\end{split}
\end{equation}

\noindent where $\hat{H}_{\ce{el}}$ is the potential part of the Hamiltonian defined in Equation \ref{eq:vibronic_coupling_ansatz}. We note that Equation \ref{eq:NuclearGrad} holds for all normal modes except for $Q_{10}$, for which $V_0$ has a non-zero gradient at $Q=0$ (Section S1 of the Supporting Information). The first two terms on the right-hand side of Equation \ref{eq:NuclearGrad} represent the gradients of the individual electronic states, referred to as the intra-state gradients, weighted by the state populations. The last term, which is directed along the derivative coupling vector and termed the inter-state gradient, stems from interference between the wave packet components on the two electronic states. This last term is thus the ``attochemical" component of the gradient, and the following discussion focuses on the nuclear motion that it induces. 
The symmetry selection rule for $\lambda_{\alpha}^{(i,j)}$ in Equation \ref{eq:Symmetry}b dictates that 
the inter-state gradient is directed along the torsional mode $Q_4$ for the $\tilde{\ce{X}}$+$\tilde{\ce{A}}$ coherent superposition, and has components along modes $Q_5$ and $Q_{11}$ for the $\tilde{\ce{A}}$+$\tilde{\ce{B}}$ superposition.
The displacement vectors for the three modes are shown in Figure \ref{fig:Fig0}.

\begin{figure*}[h!]
    \centering
     \includegraphics[]{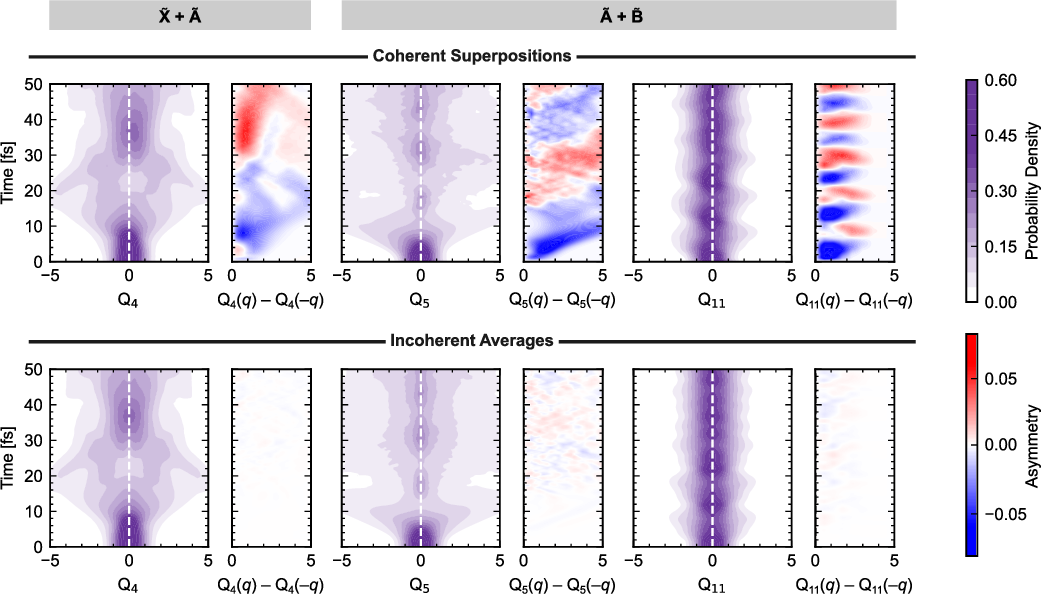}
    \caption{Top: Time evolutions of reduced nuclear densities and nuclear density asymmetries following excitations of the $\tilde{\ce{X}}$+$\tilde{\ce{A}}$ (left) and the $\tilde{\ce{A}}$+$\tilde{\ce{B}}$ (right) equally weighted in-phase coherent superpositions. The results are shown along the normal modes that exhibit non-zero inter-state gradients for each superposition. 
    Bottom: Same as the top panel, but for the incoherent averages of the dynamics observed following excitations to the relevant individual electronic states. 
    }
    \label{fig:NuclearDensities}
\end{figure*}

The time evolutions of the nuclear densities along $Q_4$ following excitation of the $\tilde{\ce{X}}$+$\tilde{\ce{A}}$ coherent superposition, and along $Q_{5}$ and $Q_{11}$ following excitation of the $\tilde{\ce{A}}$+$\tilde{\ce{B}}$ superposition, are shown in the top panels of Figure \ref{fig:NuclearDensities}. The nuclear densities evolve asymmetrically along these normal modes, as highlighted by the plots of $Q_{\alpha}(q)-Q_{\alpha}(-q)$. By the symmetry selection rule for $\kappa_{\alpha}^{(i)}$ in Equation \ref{eq:Symmetry}a, the intra-state gradients are zero along $Q_{4}$, $Q_{5}$, and $Q_{11}$,
suggesting that the inter-state gradients are responsible for this behavior. 
This is confirmed by the absence of asymmetry for the incoherent averages presented in the bottom panels of Figure \ref{fig:NuclearDensities}. Notably, the nuclear density asymmetries, which thus stem from the coherences between the initially populated electronic states, persist throughout the 50\,fs simulation time--despite the much faster timescale of electronic decoherence. 

\begin{figure}[h!]
    \centering
    \includegraphics[]{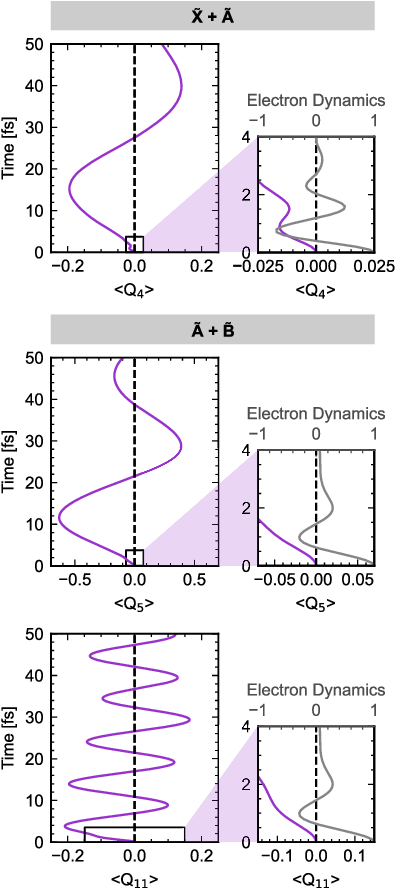}
    \caption{Time evolutions of nuclear position expectation values following excitations of the $\tilde{\ce{X}}$+$\tilde{\ce{A}}$ (top) and $\tilde{\ce{A}}$+$\tilde{\ce{B}}$ (bottom) equally weighted in-phase coherent superpositions. The results are shown along the normal modes that exhibit non-zero inter-state gradients for each superposition. The insets show magnifications of the early-time position expectation values, along with the electron dynamics, defined as $\rho_{ij}+\rho_{ji}$. Note that the x-axis ranges differ between the normal modes. 
}
    \label{fig:ExpectationValues}
\end{figure}

The nuclear density asymmetries in Figure \ref{fig:NuclearDensities} exhibit sinusoidal temporal modulations. To investigate the nature of this oscillatory behaviour, Figure \ref{fig:ExpectationValues} reports the position expectation values along $Q_{4}$, $Q_{5}$ and $Q_{11}$. We address first the early-time nuclear dynamics that take place before electronic decoherence. These are shown in the insets of Figure \ref{fig:ExpectationValues}. The insets also display the electron dynamics, represented by the sum of the off-diagonal elements of the reduced density matrix ($\rho_{ij}+\rho_{ji}$). The electron dynamics oscillate between 1 and -1 as the hole migrates across the molecular backbone. The oscillations are damped rapidly due to decoherence. The periods of the oscillations align well with those predicted from the energy gaps between the electronic state pairs at the Franck-Condon point: $h/(E_{\tilde{\ce{A}}}-E_{\tilde{\ce{X}}})=1.45$\,fs and $h/(E_{\tilde{\ce{B}}}-E_{\tilde{\ce{A}}})=2.54$\,fs for the $\tilde{\ce{X}}$+$\tilde{\ce{A}}$ and $\tilde{\ce{A}}$+$\tilde{\ce{B}}$ superpositions, respectively.

\noindent The electron dynamics produce instantaneous inter-state nuclear gradients, leading to small-amplitude oscillations in the nuclear position expectation values at the same frequencies. This effect is most clearly seen for the $\tilde{\ce{X}}$+$\tilde{\ce{A}}$ superposition. The signs of these oscillations can be reversed in a predictable (Equation \ref{eq:NuclearGrad}) and well-documented\cite{Meisner_Vacher_Bearpark_Robb_2015, Vacher_Albertani_Jenkins_Polyak_Bearpark_Robb_2016, Arnold_Vendrell_Welsch_Santra_2018} fashion by changing the relative phase between the electronic states comprising the superpositions from $\varphi=$0 to $\pi$ (Section S4 of the Supporting Information). 

The small-amplitude, short-period oscillations in the position expectation values observed before electronic decoherence are superimposed on larger-amplitude, slower oscillations that persist throughout the simulations (main panels of Figure \ref{fig:ExpectationValues}). Damped oscillations in a nuclear coordinate, driven by electron dynamics, overlaid on a slower evolution, have previously been reported in on-the-fly Ehrenfest simulations of modified bismethylene-adamantane\cite{Vacher_Albertani_Jenkins_Polyak_Bearpark_Robb_2016}, though the nature and lifetime of the slow evolution were not addressed in that study. More broadly, to our knowledge, the impact of initial electronic coherences in a polyatomic molecule on the nuclear dynamics \textit{after} decoherence has not been addressed theoretically in the literature. Upon excitation of the $\tilde{\ce{A}}$+$\tilde{\ce{B}}$ superposition, the periods of the slow oscillations that persist beyond electronic decoherence differ along $Q_5$ and $Q_{11}$, suggesting that they are characteristic of the vibrational normal modes rather than of the electron dynamics. Indeed, the periods observed in the main panels in Figure \ref{fig:ExpectationValues} agree well with those of the respective vibrational normal modes (Section S5 of the Supporting Information). As is the case for the small-amplitude, short-period oscillations driven by the electron dynamics, the signs of the larger-amplitude oscillations with frequencies characteristic of the vibrational normal modes can be reversed by changing the relative phase between the electronic states at $t=0$\,fs from $\varphi=0$ to $\pi$ (Section S4 of the Supporting Information). This, together with the absence of oscillations for the incoherent averages (bottom panels in Figure \ref{fig:NuclearDensities}, and Figures S10 and S11 of the Supporting Information), demonstrates that the long-lived nuclear oscillations stem from the short-lived coherences between the initially populated electronic states.

The theoretical findings reported herein share similarities with recent experimental results reported by \citeauthor{Schwickert_Ruberti_Kolorenč_Usenko_Przystawik_Baev_Baev_Braune_Bocklage_Czwalinna_et_al._2022} and their associated interpretation.\cite{Schwickert_Ruberti_Kolorenč_Usenko_Przystawik_Baev_Baev_Braune_Bocklage_Czwalinna_et_al._2022, Schwickert_Przystawik_Diaman_Kip_Marangos_Laarmann_2024}. In their work, a coupling of electronic to vibronic coherences was proposed to explain sinusoidal temporal modulations with shifting periods in time-resolved spectra of glycine cation.

To conclude, the simulations presented in this letter predict that short-lived electronic coherences can induce vibrational coherences with significantly longer lifetimes. For ethylene cation, considered herein as a prototype case, the vibrational coherences persist throughout the 50\,fs simulation time. Beyond this time frame, ethylene cation dissociates via \ce{H}- and \ce{H2}-loss.\cite{Fransén_Tran_Nandi_Vacher_2024} The next step is to investigate whether the yields of these dissociation channels can be controlled through rationally designed coherent superpositions. Such control of bond breaking represents the true goal of attochemistry. 
Theoretically, addressing this challenge would require either the development of a more sophisticated model Hamiltonian capable of describing large-amplitude nuclear motions or the use of an on-the-fly quantum dynamics method such as DD-vMCG.\cite{Lasorne_Robb_Worth_2007} We stress, however, that whether short-lived electronic coherences can leave sufficiently large imprints on the nuclear motion to affect the chemical outcome remains an open question. An interesting direction for future research could be to systematically investigate how to maximize the amplitude of this imprint. We hypothesize that, in addition to ensuring roughly equal weights of the coherently populated states and strong inter-state couplings close to the Franck-Condon point (see Equation \ref{eq:NuclearGrad}), the relative duration of the coherence compared to the period of the electron dynamics may be a key factor. Specifically, it may be ideal if the electron dynamics are damped by decoherence before they reverse direction. Moreover, a relatively long electron dynamics period (i.e., a small energy gap at the Franck–Condon point) may be advantageous, as this allows the force on the nuclei to act in a single direction for a longer time. 

\begin{acknowledgement}

This work received financial support under the EUR LUMOMAT project and the Investments for the Future program ANR-18-EURE-0012 (M.V. and L.F.) L.F. acknowledges thesis funding from the Région Pays de la Loire and Nantes University. The project is also partly funded by the European Union through ERC Grant No. 101040356 (M. V.). The views and opinions expressed are however those of the authors only and do not necessarily reflect those of the European Union or the European Research Council Executive Agency. Neither the European Union nor the granting authority can be held responsible for them. S.G. also thanks the EPSRC under the COSMOS programme grant (EP/X026973/1). The simulations in this work were performed using HPC resources from CCIPL (Le centre de calcul intensif des Pays de la Loire) and from GENCI-IDRIS (Grant 2021-101353). 

\end{acknowledgement}

\begin{suppinfo}
Details on the vibronic coupling model Hamiltonian, computational details for (ML-)MCTDH simulations, validation and limitations of the model Hamiltonian, position expectation values along all normal modes for in-phase and out-of-phase superpositions, rationalization of the frequencies of the nuclear motion asymmetries, effect of including bilinear on-diagonal coupling constants, sensitivity to the inter-state coupling strength, computational details and results for three-state superpositions.  


\end{suppinfo}

\bibliography{bibliography}

\end{document}